\documentclass[10pt,conference,letterpaper]{IEEEtran}
\usepackage{times,amsmath,amssymb,epsfig}
\usepackage{multirow}
\usepackage[linesnumbered,ruled]{algorithm2e}
\usepackage[table,xcdraw]{xcolor}
\usepackage{url}

\newcommand{\sys}{\textsc{Vexus}\xspace}

\title{Exploration of User Groups in VEXUS}
\author{%
{Sihem Amer-Yahia$^{\dag\diamond}$, Behrooz Omidvar-Tehrani$^{\dag}$, Jo\~ao Comba$^{\ddag}$, Viviane Moreira$^{\ddag}$, Fabian Colque Zegarra$^{\ddag}$}%
\vspace{1.6mm}\\
\fontsize{10}{10}\selectfont\itshape
$^{\dag}$Univ. Grenoble Alpes (France), $^{\diamond}$CNRS (France), $^{\ddag}$UFRGS (Brazil)\\
\fontsize{9}{9}\selectfont\ttfamily\upshape
$^{\dag}$firstname.lastname@univ-grenoble-alpes.fr, $^{\ddag}$\{comba,viviane,fczegarra\}@inf.ufrgs.br
\fontsize{10}{10}\selectfont\rmfamily\itshape
\fontsize{9}{9}\selectfont\ttfamily\upshape
}

\begin{document}
\maketitle
\begin{abstract} 
We introduce \sys, an interactive visualization framework for exploring {\em user data} to fulfill tasks such as finding a set of experts, forming discussion groups and analyzing collective behaviors. User data is characterized by a combination of demographics like age and occupation, and actions such as rating a movie, writing a paper, following a medical treatment or buying groceries.
The ubiquity of user data requires tools that help explorers, be they specialists or novice users, acquire new insights.
\sys lets explorers interact with user data via visual primitives and builds an exploration profile to recommend the next exploration steps. \sys combines state-of-the-art visualization techniques with appropriate indexing of user data to provide fast and relevant exploration.
\end{abstract}

%
\section{Introduction}
\label{sec:intro}
Today, user data can be easily acquired from various domains, ranging from the social Web to medical records, scientific publications, and retail store receipts. This data is characterized by a combination of demographics (e.g., age, gender, occupation) and actions (e.g., rating a movie, publishing a paper, following a medical treatment). Explorers in their role as {\em data scientists}, rely on user data to conduct large-scale population studies and to gain insights on the preferences of different population segments. Explorers, in their role as {\em information consumers}, use the social Web for routine tasks such as choosing a restaurant. 

In this paper, we introduce \sys ({\bf V}isualizing and {\bf EX}ploring {\bf U}ser Group{\bf S}), a visualization framework which lets explorers understand user data. Explorers can discover and visualize groups of similar users along different dimensions of their choice. They can navigate in the group space and obtain more details on demand. \sys helps them complete two kinds of tasks: reach a single group of interest (e.g., finding an audience group for targeted advertisement) or collect users among different groups (e.g., forming a geographically diverse set of experts for a conference committee). 

User data exploration is a challenging task for two main reasons. First, its large size is daunting. For instance, {\sc BookCrossing}, a book rating dataset, contains one million ratings of 278,858 users for 271,379 books. Second, explorers often have  a partial understanding of the data and their needs.
To tackle this challenge, we build \sys on two principles: {\em aggregated analytics} \cite{sofia16,DBLP:journals/pvldb/DasADY11} and {\em interactivity} \cite{DBLP:conf/cikm/TehraniAT15}. The former suggests a group-based analysis which addresses noise and sparsity
and enables new findings in a more granular space. The latter is in opposition to a more traditional single-shot analysis, and provides means to iteratively refine one's understanding.

\vspace{1pt}
\noindent {\bf Aggregated Analytics.} The aggregation of users' demographics and actions forms groups such as {\em ``young professionals in Paris''} and {\em ``female teenagers who watch romantic movies''}. All group members share common demographics and actions that describe the group. Group-based analytics provides a summarized view of user data which facilitates navigating in the space. Once landed on a group of interest, the explorer can switch to a user-level investigation. In \sys, we build upon group-based principles that we developed in \cite{DBLP:conf/cikm/TehraniAT15}.

An explorer may be interested in a group as a whole. For instance, the group of {\em ``middle-age users in France who read thrillers''} in its entirety, is a good candidate for John Grisham\footnote{\it An American writer best known for his popular legal thrillers} books. An explorer may also need to delve into a group to choose some members. For instance, program committee chairs look for an internationally-diverse and gender-balanced committee, and hence need to choose members from various groups.

Group-based analysis can be very expensive. The number of possible groups is potentially very large as it is exponential in the number of users' demographics and actions. Any set of users with at least one demographic or action in common can form a group. As an example, with only four demographic attributes and five values for each, the number of user groups will be in the order of $10^6$. Consequently, we also need to address scalability in \sys. 

\vspace{1pt}
\noindent {\bf Interactivity.} We advocate an interactive human-in-the-loop approach where an explorer who is interested in a group of users, may request to obtain other interesting groups that are semantically connected to that group in the next iteration. We distinguish an interactive process from a random walk in the space of user groups by respecting the following key principles.

\vspace{1pt}
\noindent {\em P1: Limited Options.} The explorer must be able to see different groups without being overwhelmed (i.e., Occam's razor principle). 

\vspace{1pt}
\noindent {\em P2: Optimality.} Groups offered to the explorer must be of high quality. In other words, interactive steps must collectively optimize a quality function. This ensures the purposefulness of the interactive process and prevents statistically false local discoveries such as Simpson's paradox \cite{liu2011towards}.

\vspace{1pt}
\noindent {\em P3: Efficiency.} The train of thought of the explorer must not be lost. Hence each interactive step must be fast to preserve fluidity.

\vspace{1pt}
The following example shows how the proposed exploration principles enable the efficient analysis of user data. Consider Tiffany who wants to find a person she met at last night's party in Westford, Massachusetts (MA). She does not remember his name or any other indicating contact. Hence no querying mechanism is of help. Tiffany uses \sys to inspect the list of Mike's friends. Mike is the party host and \sys forms groups from Mike's friends (\textbf{aggregated analytics}). \sys returns three groups (\textbf{limited options}) which are {\em ``engineers in MA who work in NextWorth company''}, {\em ``engineers in bioinformatics''} and {\em ``part-time market managers in Boston''}. Those groups are diverse to provide different analysis directions and cover most of Mike's friends (\textbf{optimality}). Tiffany remembers that the person she is looking for was talking about ``data visualization'', thus he should not be working for NextWorth, a recycling company. She also remembers that he mentioned he is a full-time employee; thus he should not belong to the last group either. So she selects the group of engineers in bioinformatics. 
In the next iteration, she immediately receives three subsets of that group (\textbf{efficiency}). She notices a group of {\em ``software engineers in BioView''} (a company for cell imaging and analysis) where she finds the person she was looking for.

\vspace{3pt}
Our contributions in \sys are as follows.

\vspace{1pt}
\noindent {\bf Exploratory Analysis.} \sys is particularly targeted to scenarios of exploratory analysis where explorers have a partial understanding of the underlying user data and need to refine their objectives as they discover new insights. \sys exploits appropriate indexing paradigms to enable fluid interactions. Moreover, \sys builds an explorer profile and uses it to anticipate follow-up steps and select groups on-the-fly depending on the explorer's evolving needs. \sys serves essential applications on a variety of datasets, such as building a program committee for a conference~\cite{DBLP:conf/cikm/TehraniAT15}, assembling a team of experts in crowd data sourcing~\cite{DBLP:conf/hcomp/ValeriEA15}, 
recommending items to a group~\cite{DBLP:conf/edbt/Amer-YahiaTRS15}, and validating hypotheses such as {\em ``young professionals are more inclined to buying organic food''}~\cite{DBLP:conf/dsaa/MishraLA15}. To the best of our knowledge, \sys is the first framework which enables fluid navigation of user groups in an exploratory context.

\vspace{1pt}
\noindent {\bf Visualization.}
\sys uses state-of-the-art visualization techniques to interact with the explorer. User groups are visualized in a {\em directed force layout} to prevent clutter. Histograms and charts show detailed statistics about groups. Those statistics are displayed in {\em coordinated views} where a brush on one (e.g., histogram) updates all other statistics instantaneously. The explorer can also request to see members of a group where a {\em two-dimensional projection} provides a clustered view of those users.

\vspace{-3pt}
\section{System Architecture}
\label{sec:archi}
\begin{figure}[t]
  \centerline{\includegraphics[width=\columnwidth, keepaspectratio =
      true]{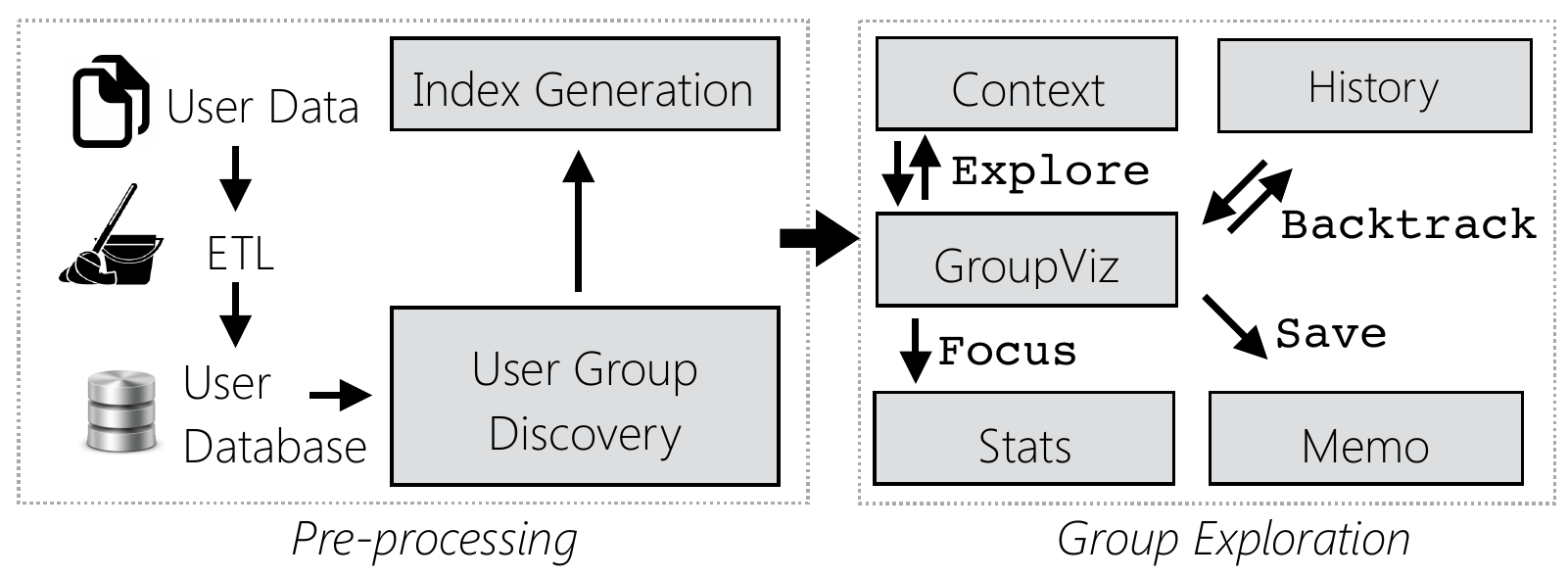}}
\caption{\label{fig:archi} \sys Architecture.}
\vspace{-20pt}
\end{figure}

Fig.~\ref{fig:archi} shows the overall architecture of our system. First, \sys pre-processes user data offline to obtain user groups. Groups form a disconnected undirected graph ${\mathcal G}$ where an edge exists between two groups if they are not disjoint. Group exploration is a navigation in that graph.

\vspace{-3pt}
\subsection{VEXUS Modules}
\noindent {\bf Pre-processing.} In the offline process, \sys receives the input user data either as a dataset (in the form of a CSV file) or as a data stream. An ETL process (including data cleaning) precedes the data import to prepare data for analysis. 
Each record in user data describes one user action (e.g., rating a book). 
We consider the generic schema [{\tt user}, {\tt item}, {\tt value}] for user data. For instance the tuple [{\tt Mary}, {\tt Mr Miracle}, {\tt 4}] means that the user ``Mary'' rates the book ``Mr Miracle'' with the score 4 (out of 5). Each user is also associated to a set of demographics.

The user data is given as input to a {\em group discovery} algorithm. 
\sys is independent of this process. For user datasets, different group discovery algorithms such as {\sc LCM}~\cite{uno2003lcm} and $\alpha$-MOMRI~\cite{DBLP:conf/pkdd/TehraniADT16} can be used. In case of user data streams,
{\sc StreamMining} \cite{jin2005algorithm} and {\sc Birch} \cite{zhang1996birch} can be employed.
For each group, its members and their common attributes will be returned.

For efficient navigation in the space of groups, we build an inverted index per group in $g \in {\mathcal G}$ that contains all groups in ${\mathcal G} - \{g\}$ in decreasing order of their similarity to $g$. We use the {\em Jaccard} distance to compute the similarity between each pair of groups. 
To reduce both time and space complexity, we only materialize 10\% of each inverted index which is shown in~\cite{DBLP:conf/cikm/TehraniAT15} to be adequate to deliver satisfying results.

\vspace{1pt}
\noindent{\bf Group Exploration.} Fig. \ref{fig:screenshot} shows a screen-shot of group exploration. 
Inspired from OLAP and in conformance with visual analytics principles \cite{bender2000functional}, we consider five visual modules in \sys: {\sc GroupViz}, {\sc Context}, {\sc Stats}, {\sc History} and {\sc Memo}. In {\sc GroupViz}, an explorer examines a limited number of groups to obtain one or more groups of interest. She can then ask to navigate to other groups which are similar to what she has already liked (i.e., interactivity). The explorer preference, captured in the form of feedback, is illustrated in {\sc Context}. The sequence of selected groups is visualized in {\sc History}. The explorer can backtrack to any previous step in {\sc History}. The explorer may request to delve into more details and observe group members. In this case, an exhaustive set of statistics will be shown in {\sc Stats}. At any stage of the process, the explorer can bookmark a group or a user in {\sc Memo}. The analysis ends when the explorer is satisfied with her collection in {\sc Memo}, which serves as her analysis goal.

{\sc GroupViz} visualizes $k$ groups in the form of circles. It is shown in previous research~\cite{miller1956human} that $k \leq 7$ is an ideal match for human perception capacity. The position of circles is enforced by a directed force layout to prevent visual clutter. The size of circles reflects the number of users in groups. Circles are be color-coded by any attribute of choice (e.g., by gender in Fig. \ref{fig:screenshot}) to provide immediate insights. The group description
is shown by hovering over the circle to provide an {\em explanation} of the group's content.

\vspace{-3pt}
\subsection{VEXUS Features}
In the following, we highlight key distinctive functionalities of \sys.

\vspace{1pt}
\noindent {\bf Interactivity.} At any given point, the explorer can click on a group $g$ in {\sc GroupViz}. Then, \sys decides which $k$ groups (conforming with principle {\bf P1}) to explore next for $g$ based on implicit feedback so far (reflected in {\sc Context}). To comply with the optimality principle {\bf P2}, the quality of $k$ groups should be verified. We consider {\em diversity} and {\em coverage} as quality objectives in \sys. Optimizing diversity provides various analysis directions and reduces redundancy in returned groups. Optimizing coverage ensures that the most interesting records appear in at least one group in the output. The results of our user study in \cite{tehrani2013towards} shows that highly diverse and covering groups are preferred as they contain informative and representative users. We use a best-effort greedy approach that we developed in~\cite{DBLP:conf/cikm/TehraniAT15} to return a local diverse and covering set of $k$ groups with a lower-bound on similarity.


Note that while all interactions in \sys occur in ${\mathcal O}(1)$, the bottleneck of the framework is the greedy process. To comply with the efficiency principle {\bf P3}, we set a time limit for the greedy process. The higher this limit, the more optimized the set of groups. We safely set the time limit to $100ms$ (i.e., continuity preserving latency \cite{fekete2016progressive}) which enables \sys to reach in average 90\% of diversity and 85\% of coverage.


\begin{figure*}[t]
  \centerline{\includegraphics[width=0.99\linewidth, keepaspectratio =
      true]{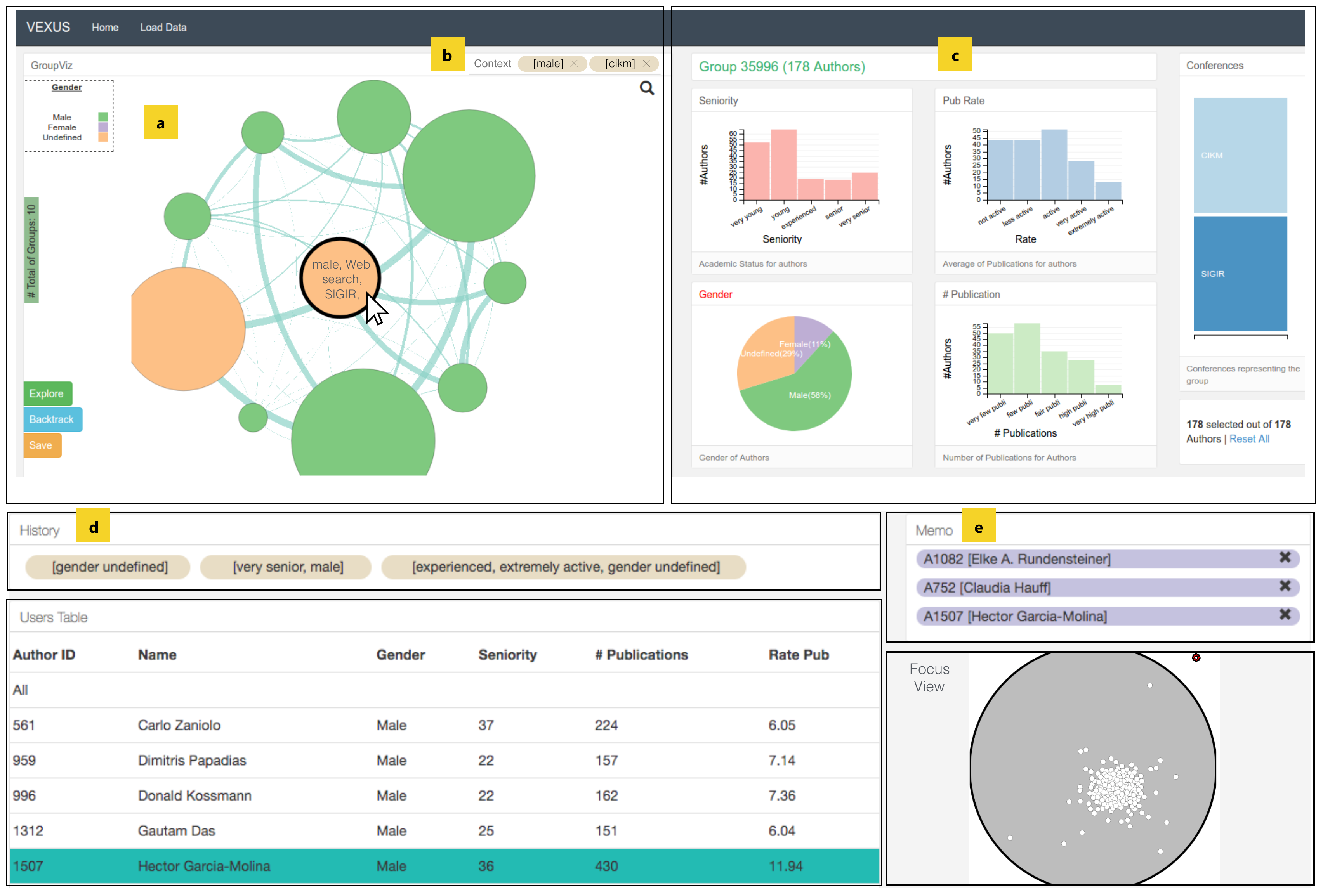}}
\caption{\label{fig:screenshot} The \sys\ Framework in Action: {\sc GroupViz} (a), {\sc Context (b)}, {\sc Stats} (c), {\sc History} (d) and  {\sc Memo} (e). A demonstration video of \sys is provided at {\em https://goo.gl/VoBC5z}.}
\vspace{-15pt}
\end{figure*}


\vspace{1pt}
\noindent {\bf Granular Analysis.} The explorer can investigate group members to inspect more details. \sys employs Linear Discriminant Analysis \cite{ji2008generalized} as a dimensionality reduction approach to obtain a 2D projection of members of a desired group (Focus View in Fig. \ref{fig:screenshot}). Members whose profile are more similar appear closer to each other. Also, histograms will show an exhaustive list of demographic distributions in {\sc Stats} module. For instance, focusing on the group of {\em ``very senior researchers in data management with a very high number of publications''} reveals that $62\%$ of its members are male. 
The explorer can brush on histograms and constrain the set of users. For instance, she can express her desire to {\em ``limit the search only to females"} by a brush on ``female'' in the gender histogram. An updated list of selected users is shown in a table. For instance, by brushing on gender to select females and on publication rate to select ``extremely active'' over the above group, the table lists {\em Elke A. Rundensteiner} (a full professor in the Computer Science department of Worcester Polytechnic Institute) with 325 publications in 26 years of her career.


\vspace{1pt}
\noindent {\bf Interoperability.} Histograms are implemented using Crossfilter charts\footnote{\it http://square.github.io/crossfilter/}. Crossfilter employs the methodology of {\em coordinated views} where a brush on one histogram updates all other statistics instantaneously. This satisfies the efficiency principle {\bf P3} at the user-level. Crossfilter's efficiency is ensured by employing the concept of {\em incremental queries} which prevents redundant query executions by sub-setting the data under the brush, on-the-fly.

\vspace{1pt}
\noindent {\bf Feedback Learning.} During the interactive process, \sys captures the explorer's feedback to personalize future steps. Feedback is considered as a probability vector over all users and demographic values. Once the explorer decides to explore a group $g$, \sys interprets this choice as a positive feedback and increases the score of $g$'s members and their common activities described in~$g$ inside the feedback vector. The vector is always kept normalized, i.e., all scores in the vector add up to $1.0$. This implicitly means that users and demographics that do not get rewarded, will gradually end up with a lower score tending to zero. 
\sys shows the explicit current status of the feedback vector in the {\sc Context} module. Hence the explorer can easily understand how \sys results are currently biased. She can easily {\em unlearn} (i.e., make \sys forget about a user or a demographic value) by deleting it from {\sc Context}. To incorporate feedback in the greedy optimizer behind the group visualizer, we consider a weighted similarity function. Intuitively, a group which is highly in line with the feedback received so far gets a higher weight, hence it is more probable to be chosen as one of the $k$ returned groups in subsequent steps. 

\vspace{-3pt}
\section{Scenarios}
We consider two user datasets: database researchers ({\sc DB-Authors}) and book ratings ({\sc BookCrossing}\footnote{\it http://www2.informatik.uni-freiburg.de/~cziegler/BX/}). Explorers can seek to achieve either a {\em single target task} (ST), where the goal is to find a single group in its entirety (e.g., finding an audience group for targeted advertisement), or a {\em multi-target task} (MT), where the goal is to identify several users of interest while exploring user groups (e.g., forming an expert-set for a conference).
In the following scenarios, we show how \sys enables explorers to achieve ST and MT tasks in an efficient and comprehensive way.

\vspace{1pt}
\noindent {\bf Scenario 1: Expert Set Formation (MT).} Our explorer can be a program committee (PC) chair whose task is to build an expert set formed by geographically distributed male and female researchers with different seniority and expertise levels. In this scenario, we employ our {\sc DB-Authors} dataset which is now available for the public.\footnote{\it https://persyval-platform.imag.fr/perscido/web/DS32/detaildataset.} \sys guides the chair in the interactive process to find colleagues to invite. The chair may start from a small group of researchers of the previous year's PC. Then \sys returns similar groups. \sys captures the feedback from the chair throughout the process and biases the exploration towards her interest. To diversify the expert set, the chair may delete a learned demographic value, e.g., ``male'', to obtain more gender-balanced results. The interactivity of \sys coupled with the feedback-based aggregations of user data makes it distinct against other expert-set formation tools such as {\sc DBPubs} \cite{baid2008dbpubs}
and {\sc Sofia} \cite{golshan2012sofia}.
Our results in \cite{DBLP:conf/cikm/TehraniAT15} show that \sys enables PC chairs to form committees of major conferences (SIGMOD, VLDB and CIKM) in less than 10 iterations on average.

\vspace{1pt}
\noindent {\bf Scenario 2: Discussion Groups (ST).} Our explorer can be an avid book reader who is looking to join an online book club. 
Having over 1,000 ratings (ranging from 1 to 10 but mostly high) for her favorite author, Debbie Macomber (author of contemporary women's fiction), the explorer navigates groups of users in {\sc BookCrossing} (as a user rating dataset) using \sys to find discussion groups. For instance, she discovers a group with whom she agrees (e.g., people who like fiction books) and another group with whom she disagrees (e.g., people who like gender-neutral books). The user study in \cite{DBLP:journals/pvldb/DasADY11} shows an 80\% satisfaction of exploring rating datasets via user groups in contrast to individuals.


\bibliographystyle{abbrv}

{\footnotesize
\bibliography{demo}}

\end{document}